\documentclass[aps,pra,showpacs,preprintnumbers,amssymb,superscriptaddress,twocolumn]{revtex4}

\usepackage{epsfig}   
\usepackage{graphicx}
\usepackage{dcolumn}
\usepackage{bm}

\begin{document}

\title{Delay of Squeezing and Entanglement using\\
Electromagnetically Induced Transparency in a Vapour Cell}

\author{G.~H\'{e}tet}
\affiliation{ARC COE for Quantum-Atom Optics, Australian National
  University, Canberra, ACT 0200, Australia}

\author{B.~C.~Buchler}
\affiliation{ARC COE for Quantum-Atom Optics, Australian National
  University, Canberra, ACT 0200, Australia}

\author{O.~Gl\"ockl}
\affiliation{ARC COE for Quantum-Atom Optics, Australian National
  University, Canberra, ACT 0200, Australia}

\author{M.~T.~L.~Hsu}
\affiliation{ARC COE for Quantum-Atom Optics, Australian National
  University, Canberra, ACT 0200, Australia}
  
\author{A.~M.~Akulshin}
\affiliation{Centre for Atom Optics and Ultrafast Spectroscopy and ARC COE for Quantum-Atom Optics, Swinburne University of Technology, Melbourne, 3122, Australia}

\author{H.~-A.~Bachor}
\affiliation{ARC COE for Quantum-Atom Optics, Australian National
  University, Canberra, ACT 0200, Australia}

\author{P.~K.~Lam}
\email[Email: ]{ping.lam@anu.edu.au}
\affiliation{ARC COE for Quantum-Atom Optics, Australian National
  University, Canberra, ACT 0200, Australia}

\begin{abstract}
We demonstrate experimentally the delay of squeezed light and entanglement using Electromagnetically Induced Transparency (EIT) in a rubidium vapour cell.  We perform quadrature amplitude measurements of the probe field and find no appreciable excess noise from the EIT process.  From an input squeezing of 3.1~dB at low sideband frequencies, we observed the survival of 2~dB of squeezing at the EIT output.  By splitting the squeezed light on a beam-splitter, we generated biased entanglement between two beams.  We transmit one of the entangled beams through the EIT cell and correlate the quantum statistics of this beam with its entangled counterpart.  We experimentally observed a 2 $\mu$s delay of the biased entanglement and obtained a preserved degree of wavefunction inseparability of $0.71$, below the unity value for separable states. 
\end{abstract}
\maketitle

A device that can faithfully store quantum states is becoming a necessity for many quantum information applications.  In particular, quantum memory will be a key component of quantum repeaters \cite{qr,dlcz} that can enable long distance distribution of secret keys in quantum cryptography.  Quantum memory may also facilitate process synchronization in quantum information protocols \cite{klm}.

Electromagnetically induced transparency (EIT) was proposed as a means for controlled atomic storage of quantum states of light by Fleischhauer and Lukin \cite{fleischhauer}.  In EIT, a strong control field can reversibly map and retrieve the information encoded on a weak probe field using long lived atomic ground states.  Shortly after the initial demonstration of ultraslow pulse propagation \cite{hau} and storage of classical light using EIT \cite{liu,phillips}, the same techniques were extended to single photons \cite{eisa05, chan05} and squeezed light \cite{Akamatsu1,ar,Lvovsky2, Kozuma2}.   Whilst being impressive demonstrations of the potential of EIT as a quantum memory for light, these proof of principle experiments still suffer from residual decoherence effects that limit the delay time and the efficiency.  In Ref \cite{ar}, near complete transmission of 1.6~dB of squeezing was observed, although no delay was measured.  In Ref \cite{Akamatsu1}, it was shown that starting with 1.1~dB of squeezing, about 0.2~dB survived propagation through an EIT medium delaying light by 3.1~$\mu$s.  Appel \emph{et al.} \cite{Lvovsky2} and Honda \emph{et al.} \cite{Kozuma2} reported the storage of squeezed light using EIT in a vapour cell and a magneto-optical trap (MOT), respectively.   With a storage time of 1~$\mu s$ in gas cell,  0.21~dB of squeezing was retrieved from an input of 1.86~dB \cite{Lvovsky2}.  With 1.2~dB of input squeezing, 0.07~dB was recalled from a MOT after being stored for 3~$\mu$s \cite{Kozuma2}. 

\begin{figure}[!b]
  \centering
  \includegraphics[width=\columnwidth]{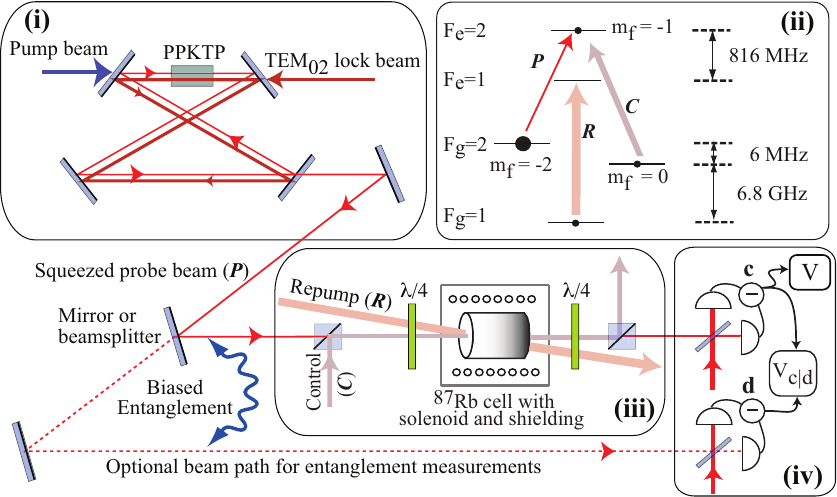}
 \caption{ Schematic of the experiment. (i) Bow-tie PPKTP optical parametric amplifier. See \cite{Hetet} for details.  The squeezed beam ($P$) is either injected directly into the EIT setup or divided using a beam-splitter to produce a pair of biased entangled beams.  (ii) The EIT level scheme.  A strong control field ($C$) pumps most of the atoms in the $m_F=-2$ state and provides the transparency for the squeezed vacuum. A repumping beam ($R$) brings atoms from the $F_g=1$ hyperfine sub-level to the $F_g=2$ level from level $F_e=1$. (iii) The gas cell used for EIT. (iv)  Joint measurements are performed using two homodyne detectors to analyse the quadrature amplitude correlations.  }
  \label{setup}
 \end{figure}
 
In this paper, we present results obtained from two experiments performed to investigate the transmission of quantum correlations through an EIT medium.  In the first experiment, we investigate the direct transmission of a squeezed light beam through the EIT medium.  With an input squeezing of 3.1~dB, we demonstrate the transmission of 2.0~dB of squeezing through an EIT feature created in a $^{87}$Rb cell filled with buffer gas.  In our second experiment, we demonstrated the delay and preservation of continuous variable entanglement by transmission through the EIT medium.  Our scheme for delaying entanglement is shown in Fig~\ref{setup}.  By splitting a single squeezed light beam, biased entanglement is generated between the two output beams of the beam-splitter \cite{biased}.  We send one of the beams through the EIT vapour cell and perform joint measurements of the quadrature amplitudes of both beams.  By analysing the quantum statistics of the joint measurements, we can directly calculate the amount of delay and entanglement between the two beams.  Delay of entanglement between remote atomic ensembles was achieved in the continuous variable regime using the off-resonant Faraday rotation \cite{juls04}.  Although the off-resonant Faraday rotation scheme can successfully store quantum optical states, the retrieval of information has to be indirectly achieved through a quantum non-demolition measurement. Entanglement delay with EIT, on the other hand, can potentially facilitate direct reversible retrieval of quantum states. 

The text is structured as follows:  in sections \ref{EITprep} and \ref{noisemeas} we discuss the preparation of an EIT feature and present measurements of noise generated by an EIT system.  In section \ref{sqzing}, we describe the construction and locking of our vacuum squeezing setup.  Squeezed vacuum propagation through an EIT feature is presented in section \ref{sqzeit}.  Lastly, we present some figures of merit for continuous variable entanglement in section \ref{entcrit} and demonstrate preservation of entanglement with an EIT induced delay in section \ref{entmeas}.

\section{EIT preparation}\label{EITprep}

In our experiments, we used the $D_{1}$ line of $^{87}$Rb (795 nm) with a level structure as shown in Fig.~\ref{setup}(ii). The atomic levels used were the $| 5^2 S_{1/2}, F_g = 2 \rangle$ for the ground state and the $| 5^2 P_{1/2}, F_e = 2 \rangle$ for the excited state. The coupling beam accessed the $| 5^2 S_{1/2}, F_g = 2, m_F = -2 \rangle$ Zeeman sublevels, and the probe beam the $| 5^2 S_{1/2}, F_g = 2, m_F = -2 \rangle$ sublevel.  Both beams were derived from a Ti:Sapphire laser (Coherent MBR). The degeneracy of the Zeeman sublevels was broken using an externally applied longitudinal magnetic field of 8.5 Gauss. To maintain the two photon resonance condition required for EIT, the control beam was frequency shifted by 6~MHz with respect to the probe light using two cascaded AOMs in a double-pass configuration. This non-degenerate configuration greatly simplifies the alignment procedure used to optimize the EIT.  When the beams are frequency degenerate, residual polarization cross-coupling between the probe and the control beams leads to parasitic low frequency fluctuations of the beam powers.  The introduced Zeeman shift between the ground states shifts these fluctuations to a frequency of 6~MHz, which is well outside of the measurement bandwidth.

The schematic of the EIT set-up is shown in Fig.~\ref{setup}(iii). The EIT medium consists of a 7.5 cm long vapour cell containing isotopically enhanced $^{87}$Rb, heated to 70$^{\circ}$C and filled with 5~Torr of Helium buffer gas. The cell is AR coated on the outside windows, which gives 92\% transmission in the absence of any active atoms.  This represents the best possible transmission our EIT system can achieve. In order to reduce stray magnetic fields, $\mu$-metal shielding was used around the cell.  The diameters of the control $(C)$ and probe $(P)$ beams were around 2~cm and 0.3~cm inside the vapour cell, respectively.  A 20~mW/cm$^2$ repump beam $(R)$ from an external cavity diode laser was used to bring atoms from the F=1 ground state hyperfine level to the F=2 ground state (as depicted in Fig.~\ref{setup}(ii)).  The repumping enhances the optical depth seen by the weak probe field without significant impact on the ground state coherence. This contrasts with an increase in cell temperature.  In this case, the optical depth increases, but so does the speed of the atoms.  The average time an atom spends in the optical beam is therefore reduced and the consequence is reduced EIT transmission.

Squeezed light is usually easier to generate at high sideband frequencies, while EIT has optimum transmission at low sideband frequencies. Our measurements of the EIT bandwidth and transmission focussed on a sideband frequency of 50~kHz, being a compromise between best squeezing and best EIT performance.  Defining the bandwidth of the EIT window to be the width-at-half-maximum, or 3dB point, we show the transmission of 50~kHz sidebands as a function of the EIT bandwidth in Fig.~\ref{bdw}(a).  The bandwidth was determined by applying a broadband 20~dB modulation and measuring the transmission on a spectrum analyser.  Due to the presence of other atomic levels and residual single-photon and two-photon detunings, the EIT window is not perfectly symmetric.  Our measurement technique averages over any such asymmetry.  The results show that a best transmission of 90\% for the 50~kHz sidebands with an EIT bandwidth of 500~kHz.  This is very close to the transmission limit of our cell of 92\%, which is shown by the grey area of Fig.~\ref{bdw}(a).
\section{Noise measurement and interpretation}\label{noisemeas}

\begin{figure}[!t]
  \centering
  \includegraphics[width=\columnwidth]{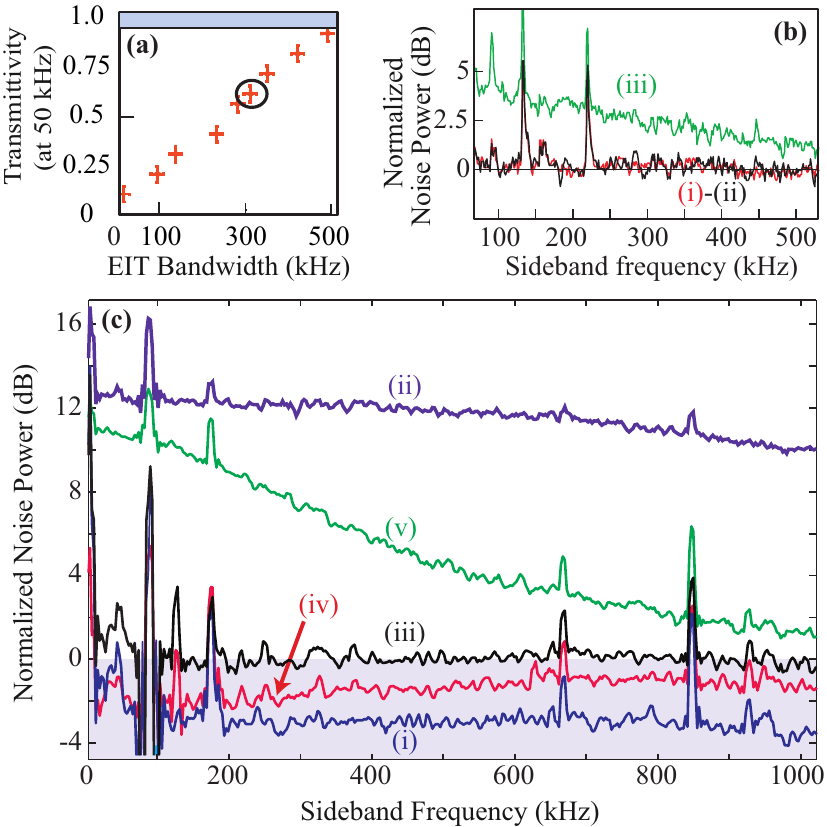}
  \caption{ (a) EIT bandwidth as a function of the transmission measured at 50 kHz, obtained by varying the control beam power. The circle is the regime where graph (c) was obtained. The cell transmission limit is shown by the grey area. (b) Noise measurements of an EIT medium in enhanced $^{87}$Rb vapour cells,  (i) Shot noise, (ii) with buffer gas and (iii) without buffer gas. (c) Transmission spectra of the squeezed light through an enhanced $^{87}$Rb vapour cell with buffer gas. (i) Squeezing and (ii) anti-squeezing measured off-resonance; (iii) shot noise;  and (iv) squeezing and (v) anti-squeezing measured on resonance under EIT conditions.}
  \label{bdw}
\end{figure}

Before sending a squeezed vacuum through the EIT system, we measured the noise introduced into the probe mode by the atoms prepared without a coherent probe beam.  In other words, we examined the properties of an  EIT system where the control field is on, but the input probe state is just a vacuum. The atomic noise measurement was made using a homodyne detector mode-matched to the probe vacuum mode. Fig~\ref{bdw}(b), trace (i) shows the shot noise level and Fig~\ref{bdw}(b)-trace (ii) shows the noise measurement made with an EIT window of 300~kHz and a vacuum state probe.  This result shows that the atoms do not add noise to the probe mode. We can reasonably expect the same behaviour for a squeezed vacuum mode which contains a few photons.

When performing the same experiment on a rubidium vapour cell without buffer gas, large excess noise was observed as can be seen Fig.~\ref{bdw}(b), trace (iii).  This noise was also measured in Ref.~\cite{hsu}. We attribute the excess-noise to non-optimum pumping into the $m_F=-2$ Zeeman sublevel and/or inelastic collisions with the cell walls. Both these mechanisms result in non-negligible atomic populations of Zeeman sublevels that are not interacting with the probe.  

Two effects are expected with residual populations in the levels $m_F=-1$ and $m_F=0$, resonant with the control field. It has been shown that this situation will give rise to gain in the probe mode \cite{HetetPRA}. This gain will in turn give excess noise following the EIT transmission window as observed in \cite{hsu}. Another possibility is that the fluorescence generated due to the pumping induced from the control field is being detected on the probe mode.  The fluorescence emitted into the probe mode will be filtered by the transmission spectrum of the EIT feature leading again to excess noise within the EIT window. 

These spurious effects are greatly reduced with the use of buffer gas or cold atoms that ensure a longer period spent within the control beam and reduce or eliminate atomic collisions with the cell wall. The ground state coherence can survive many collisions with the buffer gas which results in longer interaction time and narrower EIT features  \cite{Lvovsky4}. The situation also seems favorable in paraffin coated cells where no excess noise was observed \cite{cvi}.

\section{Squeezing preparation}\label{sqzing}

In order to achieve high transmission and slow propagation of squeezed light, a squeezed light source operating at sideband frequencies within the EIT window is required. The generation of squeezed vacuum light at the rubidium $D_1$ line was achieved recently in several laboratories using optical parametric oscillators (OPO) \cite{tanimura, Lvovsky2,Hetet}. Our squeezed light source \cite{Hetet} was shown to generate more than 5~dB of squeezed light at sideband frequencies down to 150kHz. Several modifications were made to the original set-up in order to bring the squeezing to even lower frequencies

Firstly, the homodyne detection setups were controlled using quantum noise-locking \cite{mckenzie_noiselock,schori,laurat}. This system makes use of the quadrature asymmetry of squeezed states.  The noise power over some range of sideband frequencies is measured and used to derive an error signal that will lock the homodyne detectors to the desired quadrature.
The advantage of this technique is that it does not rely on any coherent amplitude in the squeezed beam. Injection of coherent amplitude into the OPO has been identified as one of the most significant reasons for poor squeezing at low frequencies \cite{mckenzie}. In our experiment, noise in a frequency band between 0.1 and 1~MHz is used to lock the phase of the local oscillators. One issue with this technique, when applied in the context of EIT, is that at high optical depths and low control beam powers, a large part of this frequency range can be absorbed by the EIT medium.   

With noise-locking in place, squeezed light could be measured down to 20~kHz. Below this frequency, another source of noise was observed. In order to actively stabilise the OPO cavity for vacuum squeezing, we used a TEM$_{00}$ beam that travelled in the reverse direction around the OPO cavity.  It transpired that this beam was partially reflected by the surfaces of the OPO crystal leading to some residual coherent amplitude in the squeezed output. To avoid this problem a frequency shifted backwards propagating TEM$_{02}$ transverse mode was used to lock the OPO cavity, as shown in Fig. \ref{setup}(i). The combination of noise-locking and a frequency shifted OPO locking beam allowed us to produce stably locked squeezing down to 200~Hz. 

The squeezing and antisqueezing results are shown in Fig.~\ref{bdw}(c). The time domain signal was low-pass filtered at 1.9~MHz, measured at a sampling rate of 5MHz and then Fourier transformed. 
For these measurements, the homodyne visibility was 97~\% and the passive losses due to the polarising optics, the windows of the Rb vapour cell, and the detection efficiency were evaluated to be 15~\%.
When operating the OPO at a classical gain of 10 (with about 50 mW of pump light), the squeezing level, trace (i), is around 3.2~dB below the shot noise,  shown in trace (iii).  We infer an initial squeezing of 4.4 dB at the OPO output.
The anti-squeezing is about 12~dB, trace (ii).  The slight roll down of the anti-squeezing noise power is due to the OPO cavity response.

\section{Squeezed light propagation through an EIT window.}\label{sqzeit}

With this low sideband frequency squeezing and without atomic noise generated from the EIT system, we can probe the efficiency of EIT as quantum delay line.
Fig.~\ref{bdw}(c)~traces (v) and (iv) shows the typical transmission spectrum of squeezing and anti-squeezing through the EIT system. Around 2~dB of vacuum squeezed light was observed in the low frequency range. The roll up of the noise corresponds to the EIT Lorenztian transmission window. The antisqueezing displays the same feature and rolls down from 12~dB to almost 0 at higher sideband frequencies. The DC loss in this regime does not exceed 50\% and still allows us to use noise-locking to stably control quadrature detection on the output homodyne.

This experimental set-up alone however does not allow a direct measurement of the delay experienced in the EIT medium. To do this measurement, we will split the squeezed beam into two parts and compare the quantum correlations between the part that is directly detected and the other part that goes through the EIT, as shown in Fig.~\ref{setup}.
This also allows us to demonstrate the delay of continuous variable entanglement, as suggested, for example in Ref. \cite{peng}.

\section{Criteria for continuous variable Entanglement}\label{entcrit}

\begin{figure}[!b]
  \centering
  \includegraphics[width=\columnwidth]{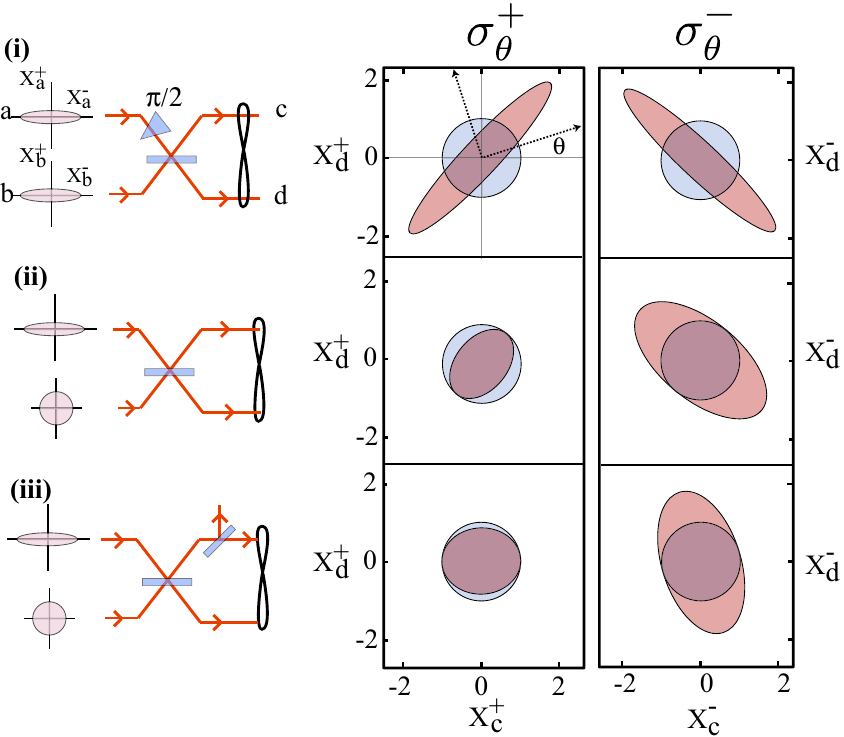}
  \caption{ Different continuous variable entangled states. i)~Two squeezed light sources $a$ and $b$ are mixed on a 50/50 beam-splitter. The resulting outputs $c$ and $d$ possess quantum correlations. On the right, the conditional deviation graphs for the output amplitude and phase quadratures. The blue circles represent the QNL conditional deviation and the red ellipses the entangled beam conditional deviation. 
 ii) Beam $b$ is replaced by a vacuum state.
 iii) Beam $b$ is a vacuum state and beam $c$ experiences some loss. Explanations in the text.
 }
  \label{entanglement}
\end{figure}

To obtain entanglement in the continuous variable regime, two identical amplitude squeezed light sources, $a$ and $b$, can be mixed on a beam-splitter with a $\pi/2$ phase difference between them \cite{Ou,bowenE}, as shown in Fig.~\ref{entanglement}(i).
A correlation measurement between the two outputs, $c$ and $d$, can be performed using two homodyne detectors measuring the uncertainty of the quadrature operators $\hat{{\rm X}}_d^{\pm}$ and $\hat{{\rm X}}_c^{\pm}$.
It can be shown that in the case of $a$ and $b$ being pure squeezed states, measuring any quadrature of $c$ will allow us to infer the corresponding quadrature of $d$ with uncertainty better than the quantum noise limit (QNL).

Formally, the variance of the conditional probability distribution of the signal $\hat{{\rm X}}_c^{\pm}$ given knowledge of the signal $\hat{{\rm X}}_d^{\pm}$ can be written as \cite{reid} 
\begin{eqnarray}
{\rm V}^{\pm}(c|d)&=&{\rm V}^{\pm}_c\big(1-\frac{|\langle \hat{{\rm X}}_c^{\pm}  \hat{{\rm X}}_d^{\pm} \rangle|^2}{{\rm V}_d^{\pm}{\rm V}_c^{\pm}}\big)
\end{eqnarray}
where $V_{c,d}^{\pm}$ is the variance of the amplitude/phase quadrature fluctuations of the beams $c$ and $d$.
When using amplitude squeezed beams as input states, the conditional variance between $c$ and $d$ will be below the QNL, given by ${\rm V}^{\pm}(c|d)=1$, indicating a non-classical correlation between them.

The right-hand side of Fig. \ref{entanglement}(i) displays the correlations between the amplitude and phase quadratures of $c$ and $d$. The perimeter of the ellipses shows $\sigma_{\theta}^{\pm}$ given  by
\begin{eqnarray}
\sigma_{\theta}^{\pm}=\sqrt{{\rm V}^{\pm}_\theta(1-(C^{\pm}_{\theta})^2)},
\label{sigma}
\end{eqnarray}
where ${\rm V}^{\pm}_\theta$ is the variance of the data projected onto axes at an angle $\theta$ and ${\rm C}^{\pm}_{\theta}=|\langle \hat{\rm X}^{\pm}_{\theta}\hat{\rm X}^{\pm}_{\theta+\pi/2}\rangle|^2/{\rm V}^{\pm}_{\theta}{\rm V}^{\pm}_{\theta+\pi/2}$ is the correlation also measured along the rotated axes.  For the situation shown in Fig. \ref{entanglement}(i), we find that $\sigma_{\theta}^{\pm}$ is an ellipse with its axis oriented at $+\pi/4$ for the amplitude quadratures and $-\pi/4$ for phase quadratures.  In Fig. \ref{entanglement}(i) these are shown in red.

The conditional variance ${\rm V}^{\pm}(c|d)$ can be found from $\sigma_{\theta}^{\pm}$ by measuring the square of the radius of the ellipse at the point where it crosses the horizontal axis, that is ${\rm V}^{\pm}(c|d)=(\sigma_{\theta=0}^{\pm})^2$. ${\rm V}^{\pm}(d|c)$ however will be found from the radius of ellipse at the points where it crosses the vertical axis, ${\rm V}^{\pm}(d|c)=(\sigma_{\theta=\pi/2}^{\pm})^2$. The  QNL is obtained by replacing the squeezed beams by vacuum states. The QNL forms circles of unity radius as shown by the perimeters of the blue circles in Fig. \ref{entanglement}(i).   For the case in this figure it is clear that ${\rm V}^{\pm}(d|c)={\rm V}^{\pm}(c|d)<1$.

When $\theta=-\pi/4$ and $\theta=\pi/4$, for the amplitude and phase quadratures respectively, the correlations lie inside the unity circle and reaches a minimum. To see what these minima mean, one can calculate $\sigma_{\theta}^{\pm}$ as a function of the rotation angle ($\theta$) and the input variances. We find
\begin{eqnarray}
(\sigma_{\theta}^{\pm})^2&=&\frac{1}{2}(1-{\rm sin}(2\theta)){\rm V}_{a}^{\pm}+\frac{1}{2}(1+{\rm sin}(2\theta)){\rm V}_{b}^{\mp}
\nonumber\\
&-&\frac{({\rm V}_{b}^{\mp}-{\rm V}_{a}^{\pm})^2 {\rm cos}^2(2\theta)}{2(1-{\rm sin}(2\theta)){\rm V}_{b}^{\mp}+2(1+{\rm sin}(2\theta)){\rm V}_{b}^{\mp}}
\end{eqnarray}
We see that $\sigma_{-\pi/4}^{+}=\sqrt{{\rm V}_{a}^{+}}$ and $\sigma_{\pi/4}^{-}=\sqrt{{\rm V}_{b}^{+}}$, which are the two initially squeezed quadratures. We also have $\sigma_{\pi/4}^{+}=\sqrt{{\rm V}_{b}^{-}}$ and $\sigma_{-\pi/4}^{-}=\sqrt{{\rm V}_{a}^{-}}$, which are the two anti-squeezed quadratures \footnote{Experimentally, the orientation of the ellipses depends on the error signal used to lock the homodyne detectors. Changing the slope of the error signal on one of the homodyne detectors will reverse the orientation of the ellipse.}.

 The correlation ellipses provide, in summary, a unified graphical representation of the conditional variances between two signals $c$ and $d$ and the variances of the original inputs $a$ and $b$.    These quantities will prove useful for the calculation of the entanglement figures of merit.

There are several criteria for measuring entanglement. We use the EPR criterion \cite{reid} and the wavefunction inseparability criterion \cite{Duan2000,simonP}. 
According to the EPR criterion, the product of the conditional variances ${\rm V}^+(c|d) {\rm V}^-(c|d) < 1$, for entangled beams. We can write the product of the conditional variances in terms of the input beams at the beam-splitter and find that 
\begin{equation}
{\rm V}^+(c|d) {\rm V}^-(c|d) = \frac{4 {\rm V}^-_a{\rm V}^-_b{\rm V}^+_a{\rm V}^+_b}{({\rm V}^-_a+{\rm V}^+_b)({\rm V}^+_a+{\rm V}^-_b)}
\end{equation}
It can be shown from this equation that entanglement can be obtained when the two input beams are pure squeezed states, i.e when ${\rm V}_{(a,b)}^{+}{\rm V}_{(a,b)}^{-}=1$, and for example ${\rm V}_a^{+}<1$ and ${\rm V}_b^{+}<1$. Strong entanglement will be obtained in the regimes of large and pure squeezing.
Entanglement can also be obtained when only one input beam is a pure squeezed state and the other input beam is vacuum (e.g. ${\rm V}_a^+ < 1 < {\rm V}_a^-$ and ${\rm V}_b^{\pm} = 1$). This situation is depicted Fig. \ref{entanglement}(ii). The state generated that way is called a biased entangled state \cite{biased} because of the asymmetry at the two output quadratures. The correlation plots indeed show that in this case one has ${\rm V}^+(c|d)=1$ and ${\rm V}^-(c|d)<1$ so the EPR inequality still holds.

When the losses on the two entangled beam are equal the conditional variances are the same whether the state is inferred from $c$ to $d$ or $d$ to $c$. When the losses are different on each arm, like in the situation depicted in Fig.~\ref{entanglement}(iii), the conditional variance of $d$ given $c$ is larger than for $c$ given $d$.
These different ways to infer are referred to as direct reconciliation and reverse reconciliation respectively \cite{Grosshans}. 
This gives rise to two numbers for EPR correlations.  This is seen graphically in the corresponding correlation plots (Fig. \ref{entanglement}(iii))  where the ellipses have both been rotated clockwise by an amount depending on the loss on the beam $c$. The difference between the ${\rm V}^{\pm}(c|d)$ and ${\rm V}^{\pm}(d|c)$ appears clearly.

Let us now refer to another criteria for continuous variable entanglement. It was introduced by Duan et al. \cite{Duan2000} and quantifies the degree of separability of the wave-function. A bipartite Gaussian entangled
state can be shown to be described by its correlation matrix \cite{Duan2000} which has the following elements

\begin{equation}
C^{ij}_{cd}=\frac{1}{2} \langle \hat{X}^{i}_{c} \hat{X}^{j}_{d} +
\hat{X}^{i}_{d} \hat{X}^{j}_{c} \rangle - \langle \hat{X}^{i}_{c}
\rangle \langle \hat{X}^{j}_{d} \rangle
\end{equation}
where $\{ i,j\} \in \{ +,- \}$.
Before the inseparability criterion can be applied, the correlation matrix has to be in standard form \textrm{II}, which can be achieved by
application of the appropriate local-linear-unitary-Bogoliubov-operations (local rotation and
squeezing operations) \cite{Duan2000}. The product form of the
degree of inseparability \cite{Bowen034} is then given by
\begin{equation}
\mathcal{I}=\frac{\sqrt{C_{I}^{+} C_{I}^{-}}}{k+1/k},
\end{equation}
where
\begin{eqnarray}
C^{\pm}_{I}\!\!\!&=& 
\!\!\!k C^{\pm\pm}_{xx} 
\! +\! 
(1/k)C^{\pm\pm}_{yy}\!\! 
-\!\! 2 | C^{\pm\pm}_{xy}|\\
k&=&\left(
\frac{C^{\pm\pm}_{yy}-1}{C^{\pm\pm}_{xx}-1}\right)^{\frac{1}{2}}.
\end{eqnarray}
$\mathcal{I}<1$ is a necessary and sufficient condition of
inseparability and therefore entanglement. 
In the case of equal losses on both arms it can be shown that the product form of the Duan criterion is equivalent to 
\begin{eqnarray}
\mathcal{I}&=&{\rm V}({\rm X}^{+}_c\pm {\rm X}^{+}_d){\rm V}({\rm X}^{-}_c\pm {\rm X}^{-}_d)<1
\end{eqnarray}
where ${\rm V}({\rm X}_c\pm {\rm X}_d)={\rm min}\langle ({\rm X}_c\pm {\rm X}_d)^2 \rangle$. 
This last quantity can be evaluated quite easily, for example,  from the conditional deviation ellipse Fig. \ref{entanglement}(i).
From the graph, we see that ${\rm V}({\rm X}^+_c\pm {\rm X}^+_d)=(\sigma_{-\pi/4}^{+})^2={\rm V}_{a}^{+}$ which is the squeezed quadrature of beam $a$.  On the other hand,  ${\rm V}({\rm X}^-_c\pm {\rm X}^-_d)=(\sigma_{\pi/4}^{-})^2={\rm V}_{b}^{+}$, which is the squeezed quadrature of beam $b$. In this situation $\mathcal{I}<1$ so the state is not separable. 
When the losses are different on both arms, local unitary transformations have to be done to the correlation matrix to express it in standard form II.  This process has a very simple graphical interpretation. In the case of unequal losses shown in Fig. \ref{entanglement}(iii), the minima of the ellipses no longer appear on the diagonals at $\theta = \pm \pi/4$. The local transformations are just used to reorient the ellipses so that the minima will again appear on the diagonals. The local transformations do not change the value of these minima, so we can always find $\mathcal{I}$ directly from the minima of the conditional deviation ellipses without local transformations.

\section{Entanglement measurement}\label{entmeas}

\begin{figure}[!t]
  \centering
  \includegraphics[width=\columnwidth]{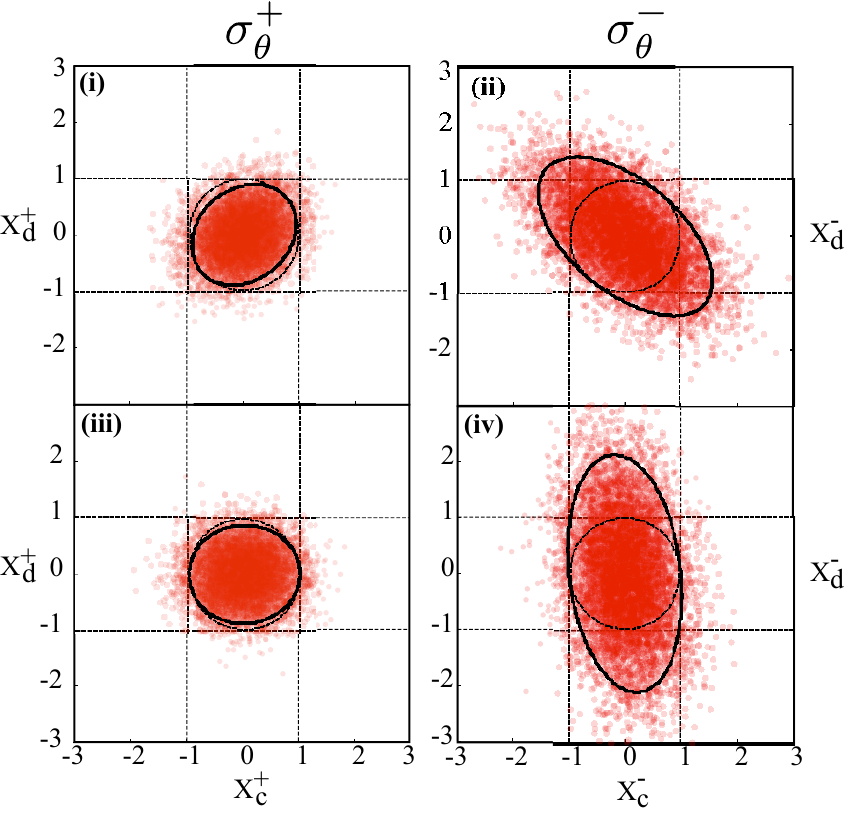}
  \caption{ Correlation measurements. (i) and (ii): Scatter plots of the amplitude and phase quadratures respectively as measured for the beams $c$ and $d$.  The lasers were not resonant and there is no EIT.
(iii) and (iv):  Data as above but with EIT switched on.   The solid black curves show the conditional deviation $\sigma_{\theta}^{\pm}$ calculated from the data.   The dashed circles show the QNL conditional deviation obtained by blocking the two entangled paths.  The coordinates of the red data points have been scaled down by a factor of two for clarity.}
  \label{biasedent}
\end{figure}

We now proceed and calculate the degree of entanglement produced by splitting our squeezed light source in two, and demonstrate that entanglement remains after transmission of one beam through the EIT medium.  This is the experiment shown in Fig.~\ref{setup}, including now the optional beam path and second homodyne detector.  In terms of the preceding discussion of entanglement criteria, this system is similar to the case of Fig.~\ref{entanglement}(iii) where the passive loss is now the EIT medium.  As we will see, biased entanglement is sufficient to find the delay and show preservation of wavefunction inseparability.  

 An initial characterization of our entanglement source was done off-resonance, i.e. without any active atoms in the gas cell.
Around 1.5~dB of squeezed light was sent though the vapour cell and the remaining 1.5~dB in free space. The visibilities on the cell and free space homodyne detectors were 97~\% and 99~\% respectively.

The subtracted signals on both homodyne detectors were acquired for 0.5$s$, mixed down digitally from 50~kHz to DC and low pass filtered at 10~kHz.  This process gives a picture of the time domain data in a bandwidth around 50~kHz.  When measuring amplitude quadratures on both homodyne detectors, we obtain the scatter plot shown in Fig.~\ref{biasedent}(i).   For phase quadrature measurements we get Fig.~\ref{biasedent}(ii).  From this data, we calculate the conditional deviation ellipses using Eq.~\ref{sigma}.  These ellipses are shown by the solid lines.  The dashed circles show the QNL.  We note that the ellipses are not rotated from the diagonal axis, which demonstrates that each beam experiences near equal loss.

From the conditional deviation curves we can read off the EPR and wavefunction inseparability criteria. 
We find EPR values of ${\rm V}^+(c|d) {\rm V}^-(c|d)=0.8\times1.6=1.28$ and ${\rm V}^+(d|c) {\rm V}^-(d|c)=0.8\times1.62=1.30$.  The inference from $d$ to $c$ gives a slightly larger EPR value due to small extra losses from the cell windows and the difference in the homodyne visibility. These values are above 1, so according to the EPR criterion there is no entanglement. This is primarily due to the impurity of our squeezed state.  Internal loss inside the OPO cavity always leads to squeezed states with non-minimum uncertainty and, as discussed, the EPR criterion is sensitive to the purity of the initial squeezing. 

Using the wavefunction inseparability criterion we find $\mathcal{I}=0.65$ which is a clearly well below unity.  So, while we can not show EPR, we easily show wavefunction inseperability.

Having established a performance benchmark using the off-resonant atoms we tuned the laser frequencies to obtain an EIT feature in our gas cell.   Measurements of the amplitude and phase quadratures were made as for the off-resonant case and the data is shown in  Fig.~\ref{biasedent}(iii) and (iv).
\begin{figure}[!b]
  \centering
  \includegraphics[width=\columnwidth]{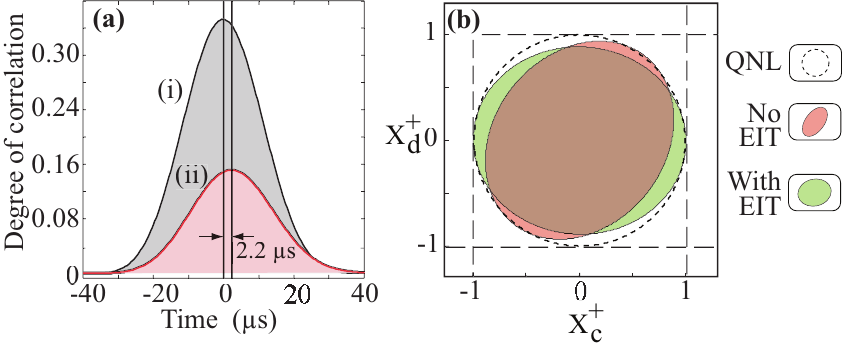}
  \caption{ a) Degree of correlation between the two entangled arms as a function of delay, $\tau$. b) Output photocurrent versus input photocurrent conditional deviations. The circle represents the shot noise limit. The dashed and plain line the off-resonance and EIT case.}
  \label{Finalres}
\end{figure}
We note that the ellipse is now rotated compared to the off resonance case, indicating the presence of loss in the EIT medium.
To better quantify this, we again use our entanglement criteria.
As expected, we find distinct EPR measures depending on how the inference is done for the conditional variance measure.
We find ${\rm V}^+(c|d) {\rm V}^-(c|d)=1\times1.25=1.25$ and ${\rm V}^+(d|c) {\rm V}^-(d|c)=0.8\times4=3.20$. 
We note that the presence of loss in the EIT medium does not change the conditional variance significantly when inferring from the beam propagating in free space. 

From the criterion for inseparability we find $\mathcal{I}=0.71$, after converting the covariance matrix to the required standard form or using ${\rm min}_{\pm} \sigma_{\theta}$. This value is higher than the off-resonance case but still below 1,
demonstrating that our EIT system preserves inseparability.

We now compute the degree of squeezing correlation $g(\tau)=\langle {\rm X}_c^{+}(t){\rm X}_d^{+}(t-\tau) \rangle$ as a function of the delay, $\tau$, between $c$ and $d$.  By looking for a peak correlation as a function of $\tau$ we can find the delay introduced by the EIT transmission. 
Fig. \ref{Finalres}(a) represents the degree of correlation between $c$ and $d$ with the atoms off (i) and on resonance (ii). This shows that EIT delayed the transmission of beam $c$ by 2.2 $\mu$s.  Some amount of correlation is clearly lost in transmission through the EIT as the peak of curve (ii) is substantially lower than case with no EIT.
Fig. \ref{Finalres}(b) compares the amplitude quadrature conditional deviations with and without EIT.  The reduced correlation is also clear in this figure.

Larger delays could not be observed due to the lack of noise-locking stability in the high optical depth or small control beam regime. Decreasing the control beam or increasing the optical depth cuts-off the frequency band necessary to obtain reliable noise-locking.  At such low frequencies however, getting long term stability is particularly crucial since the intregration times required for measurement are also larger.  An alternative to noise-locking would be some form of coherent vacuum locking, as demonstrated by Vahlbruch \emph{et al.} \cite{val}.  In this scheme a frequency shifted beam is injected into the OPO.  This beam also senses the OPO gain and can therefore be used for quadrature locking downstream.  The only complication here is that this frequency shifted beam must also pass through the gas cell without disturbing the EIT properties or being absorbed.

\section{Conclusion}
A narrow and large contrast EIT feature was generated in a warm $^{87}$Rb vapour in the presence of buffer gas. Using a buffer gas allowed us to obtain quantum noise limited delay, removing the excess noise observed previously \cite{hsu}.
Using this system we demonstrated the efficient transmission of squeezing through an EIT feature.  Out of an initial 3.2~dB of squeezing, 2~dB was observed at the EIT output.  By splitting the squeezing in two, we generated a source of biased entanglement that could be used to measure the delay due to EIT transmission and also demonstrate preservation of wavefunction inseperability.  With this method found our EIT system to have a delay of 2.2~$\mu s$.   The wavefunction inseparability after EIT delay of one half of the entangled state was measured to be $0.71$. This result is a promising step towards the reversible storage of continuous variable quantum information, a necessary milestone for many quantum information protocols. 

\section{acknowledgement}

We thank J.~Ortalo, J.~Cviklinski, N.~B.~Grosse and A.~I. Lvovsky for useful discussions.  This project is funded by the Australian Research Council's Centre of Excellence scheme.

\end{document}